\documentclass[twocolumn,amsmath,amssymb,pra,superscriptaddress]{revtex4-1}

\usepackage{mathpazo} 

\usepackage{graphicx}
\usepackage{dcolumn}
\usepackage{bm}
\usepackage{microtype}
\usepackage{helvet}

\usepackage{caption}
\DeclareCaptionLabelFormat{plain}{Figure #2 $\boldsymbol |$ }

\begin{document}

\title{Sub-thermal to super-thermal light statistics from a disordered lattice via deterministic control of excitation symmetry}

\author{H. Esat Kondakci}\email{esat@creol.ucf.edu}
\affiliation{CREOL, The College of Optics $\&$ Photonics, University of Central Florida, Orlando, FL 32816, USA}

\author{Alexander Szameit}
\affiliation{Institute of Applied Physics, Friedrich-Schiller-Universit\"{a}t Jena, 07743 Jena, Germany}

\author{Ayman F. Abouraddy}
\affiliation{CREOL, The College of Optics $\&$ Photonics, University of Central Florida, Orlando, FL 32816, USA}

\author{Demetrios N. Christodoulides} 
\affiliation{CREOL, The College of Optics $\&$ Photonics, University of Central Florida, Orlando, FL 32816, USA}

\author{Bahaa E. A. Saleh}
\affiliation{CREOL, The College of Optics $\&$ Photonics, University of Central Florida, Orlando, FL 32816, USA}


\begin{abstract}
Monochromatic coherent light traversing a disordered photonic medium evolves into a random field whose statistics is dictated by the disorder level. Here, we demonstrate experimentally that light statistics can be deterministically tuned in certain disordered lattices even when the disorder level is held fixed -- by controllably breaking the excitation-symmetry of the lattice modes. We exploit a lattice endowed with disorder-immune chiral symmetry in which the eigenmodes come in skew-symmetric pairs. If a single lattice site is excited, a `photonic thermalization gap' emerges: the realm of sub-thermal light statistics is inaccessible regardless of the  disorder level. However, by exciting two sites with a variable relative phase, as in a traditional two-path interferometer, the chiral symmetry is judiciously broken and interferometric control over the light statistics is exercised, spanning sub-thermal and super-thermal regimes. These results may help develop novel incoherent lighting sources from coherent lasers. 
\end{abstract}

\small
\maketitle

\begin{figure}[b]\centering \includegraphics[scale=1]{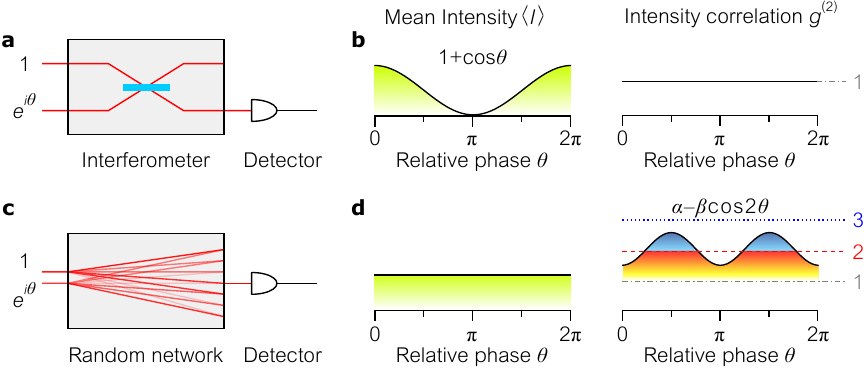} \caption{\label{Fig:Intro}  \textbf{Light-statistics interferometry in random networks.} \\ (a) Schematic of traditional two-path interferometry. Coherent fields with a relative phase $\theta$ interfere in a deterministic system (depicted here as a simple beam splitter). (b) The output intensity $I$ varies sinusoidally with $\theta$. The field remains coherent and $g^{(2)}\!=\!1$. (c) Light-statistics interferometry in a disordered system. Just as in (a), two coherent fields with relative phase $\theta$ enter the system. (d) While the ensemble averaged intensity $\langle I\rangle$ at the output is independent of $\theta$, $g^{(2)}$ varies sinusoidally with it, resulting in a photon-statistics interferogram.}\end{figure}

In optical interferometry, typically two beams are combined with a relative phase that sinusoidally modulates the intensity \cite{Born1999}. This general scenario -- fundamental to optics -- is depicted in Fig.~\ref{Fig:Intro}(a),(b). If the interferometer is replaced by a \textit{random} network with multiple input and output channels (Fig.~\ref{Fig:Intro}(c)), varying the phase between two or more coherent incident fields is not anticipated to yield interferometric control over the exiting field when an ensemble of disorder realizations is considered. Moreover, we expect that the higher-order optical statistics, such as the normalized intensity correlation $g^{(2)}=\langle I^2 \rangle/\langle I \rangle^2$, to be altogether independent of the input phases after traversing this random system; $I$ is the intensity and $\langle\cdot\rangle$ denotes ensemble averaging \cite{Goodman2000}. Here, nothing is fluctuating in time but a particular distribution of the random network is regarded as a single realization of a statistical ensemble defining a probability space. In certain cases, modulating the input field via feedback \cite{Saleh2011, Mosk2012} can help control some features of the output of a random system \cite{Vellekoop2010, Judkewitz2013, Nixon2013, Zhou2014a, Lai2015}. To date, such schemes have focused solely on the intensity and not on the higher-order intensity correlations. A wide range of applications would be served, however, by exercising facile control over the photon statistics, ranging from producing laser-driven white lighting \cite{Denault2013} to generating beams with low-spatial coherence \cite{Redding2012, Redding2015} or non-Rayleigh speckles \cite{Bromberg2014} for bio-imaging.

Here, we report on a class of random photonic networks that -- counter-intuitively -- enables \textit{deterministic interferometric control} over the light's statistics without modifying the disorder level of the network itself. Indeed, by altering the phase between two mutually coherent input beams, the \textit{higher-order correlations} of the emerging light are tailored while maintaining a fixed mean intensity (Fig.~\ref{Fig:Intro}(d)). In this scenario, varying the relative input phase results in sinusoidally modulating $g^{(2)}$ -- just as the \textit{intensity} in two-path interferometry changes sinusoidally with the phase (Fig.~\ref{Fig:Intro}(b)). This remarkable behavior is realized in a class of random media constrained by a disorder-immune symmetry known as `chiral symmetry', whereupon the eigenmodes occur in skew-symmetric pairs whose eigenvalues are equal in magnitude but opposite in sign in \textit{each} realization of the disordered ensemble \cite{Gade1991, Gade1993, Evangelou2003}. In other words, in random photonic networks exhibiting chiral symmetry, the eigenmodes appear in pairs whose members are associated with counter-rotating phasors of equal magnitudes (in the rotating frame that is common to all modes).

Surprising phenomena emerge in such  symmetry-constrained disordered lattices. Consider illuminating a single lattice site (corresponding to one input in Fig.~\ref{Fig:Intro}(c)), which guarantees that the modes in each chiral pair are excited with equal weights \cite{Lahini2011a}. Instead of the expected gradual increase in speckle contrast (quantified by $g^{(2)}$) at the output  with increasing disorder, we have recently predicted that an abrupt climb in $g^{(2)}$ to super-thermal statistics ($g^{(2)}=3$) occurs at asymptotically low amounts of disorder, followed by a gradual \textit{reduction} to thermal or pseudo-thermal statistics ($g^{(2)}=2$) upon \textit{increasing} the disorder level \cite{Kondakci2015b}. Such a random medium therefore witnesses the emergence of a \textit{photonic thermalization gap}: the range of sub-thermal statistics ($1<g^{(2)}<2$) is inaccessible to traversing light. Observation of this gap is predicated, however, on satisfying the modal excitation-symmetry condition. Breaking the excitation-symmetry, for example by illuminating two sites with a relative phase, allows for $g^{(2)}$ to be varied above and below the edge of the thermalization gap \cite{Kondakci2015b}.

We thus introduce a novel form of interferometry, \textit{light-statistics interferometry}, that is mediated by the class of disordered systems endowed with chiral symmetry. Starting with monochromatic coherent light for which $g^{(2)}\!=\!1$, coherent control at the input tunes $g^{(2)}$ at the output above and below the value $g^{(2)}\!=\!2$ upon averaging over an ensemble of realizations, thereby spanning the regimes of sub-thermal and super-thermal statistics \cite{Kondakci2015b}. In our experiments, we demonstrate this principle in a well-controlled model: a lattice of evanescently coupled identical waveguides with random couplings (so-called off-diagonal disorder) \cite{Soukoulis1981a, Szameit2010, Martin2011a}. By illuminating two neighboring waveguides with coherent light of equal amplitude and variable relative phase, we change the weights of the excited modes. While some coherent field distributions exploit the symmetry constraints to alleviate the randomization effect, others augment the fluctuations of the emerging light.

\section{Lattice model}

\begin{figure}[b!]
\centering \includegraphics[scale=1]{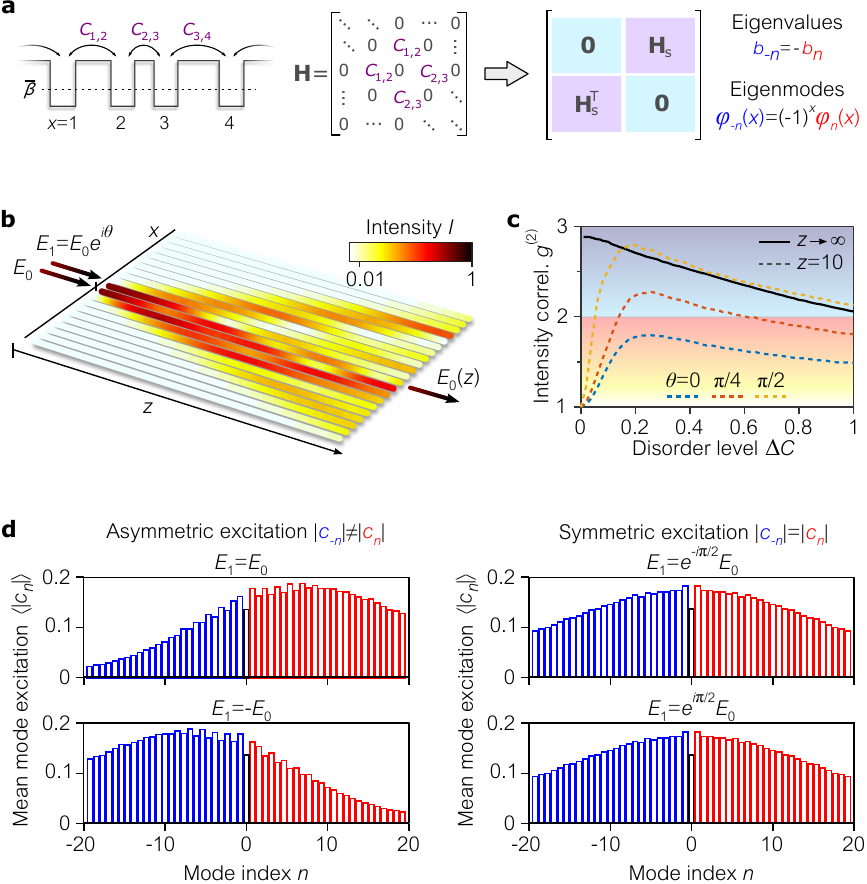}
\caption{\label{Fig:numerical} \textbf{The lattice model and photonic thermalization gap.} \\ (a) Coupled potential wells with random couplings and fixed site-energies represent an off-diagonal disordered lattice. Assuming nearest-neighbor-only coupling, the Hamiltonian $\mathbf{H}$ is a tri-diagonal matrix, which can be rearranged in block off-diagonal form in the interaction picture, a signature of chiral symmetry. (b) A waveguide array with off-diagonal disorder is coherently excited at two sites. The color along the waveguides represents the calculated intensity in a logarithmic scale. (c) Calculated  $g^{(2)}$ at $x=0$ as a function of disorder level $\Delta C$. The solid black line corresponds to single-waveguide  excitation ($x=0$) in the steady state ($z\rightarrow\infty$), while dashed lines represent $g^{(2)}$ at $z\bar{C}=10$ when two neighboring waveguides ($x=0$ and $1$) are excited, $E_1(0)=e^{i\theta} E_0(0)$, for $\theta=0$, $\pi/4$, and $\pi/2$. The ensemble size is $10^5$. (d) The mean mode-excitation amplitudes $\langle|c_n|\rangle$ are asymmetric (left) around $n=0$ for $\theta\!=\!0$ and $\pi$ and are symmetric (right) when $\theta\!=\!-\pi/2$ and $\pi/2$.}
\end{figure}

Our photonic system is modeled after the generic tight-binding lattice illustrated in Fig. \ref{Fig:numerical}(a). We consider a one-dimensional (1D) array of identical waveguides (having the same wave number $\bar{\beta}$) with nearest-neighbor-only coupling (Fig.~\ref{Fig:numerical}(b)). In matrix form, the dynamics of field propagation in the interaction picture is given by $-i\frac{d\mathbf{E}}{dz}=\mathbf{HE}$, where $\mathbf{E}=\{E_x(z)\}_{x=-N}^N$ is a vector containing the complex-field amplitudes at the lattice sites $x$ at axial position $z$, and $\mathbf{H}$ is the Hamiltonian, a coupling matrix in tri-diagonal form by virtue of nearest-neighbor-only coupling \cite{Christodoulides2003a, Schwartz2007a, Lahini2008a, Lahini2011a, Segev2013}. The field evolution can be expressed succinctly in terms of the eigenvalues $b_n$ and the orthonormal eigenmodes $\varphi_n(x)$ of $\mathbf{H}$, $\mathbf{H}\varphi_n(x)=b_n\varphi_n(x)$. If the input field is $E_{x}(0)=\sum_{n} c_{n}\varphi_{n}(x)$, where $\{c_{n}\}$ are the complex mode-excitation amplitudes, then $E_{x}(z)=\sum_{n} c_{n}\varphi_n(x)e^{ib_n z}$ and the intensity is $I_x(z)=|E_x(z)|^{2}$. We may rearrange $\mathbf{H}$ here into a block off-diagonal matrix (Fig.~\ref{Fig:numerical}(a)), which is a hallmark of chiral ensembles \cite{Gade1991, Gade1993, Evangelou2003}: it entails that the eigenvalues occur in anti-symmetric pairs $b_{-n}\!=\!-b_{n}$ and the associated eigenmodes satisfy $\varphi_{-n}(x)\!=\!(-1)^x\varphi_{n}(x)$. We consider here lattices with off-diagonal disorder having a normalized disorder level $\Delta C\!=\!W/\bar{C}$, where $W$ is the half-width of a uniform probability distribution for the random coupling coefficients of mean $\bar{C}$. Crucially, the characteristic features of chiral ensembles are disorder-immune; i.e., they hold  for each realization from a disordered ensemble \cite{Evangelou2003}.

\begin{figure*}[!t]
	\centering\includegraphics[scale=1.05]{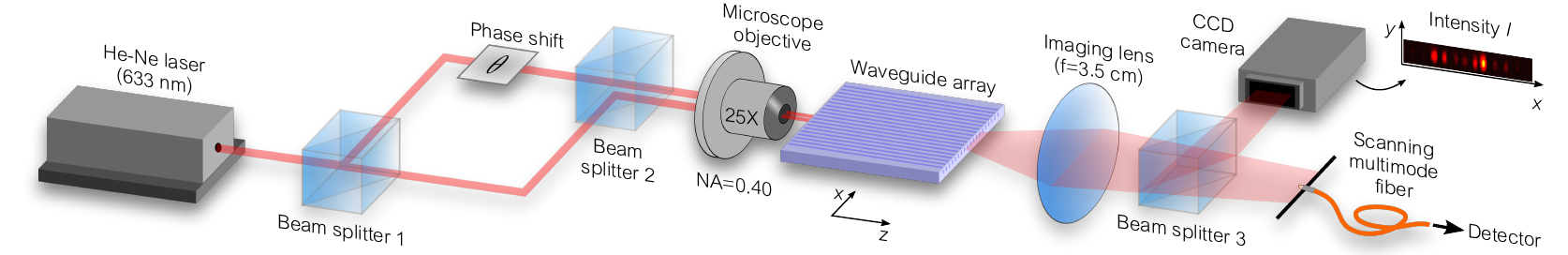}
	\caption{\label{Fig:Setup} \textbf{Experimental setup.} \\ A single-mode coherent beam from a He-Ne laser is split equally in two paths, a phase shift $\theta$ is introduced,  and the two beams are then imaged into two neighboring waveguides within an array. The different disorder realizations are produced by translating the waveguide array along $x$ and ensuring that the input beams are re-aligned for each configuration. After magnification, the waveguide-array output is imaged to a CCD camera, and a single waveguide at $x=0$ is separately imaged to a multimode fiber.}
\end{figure*}

\begin{figure*}[t]
	\centering\includegraphics[scale=1.05]{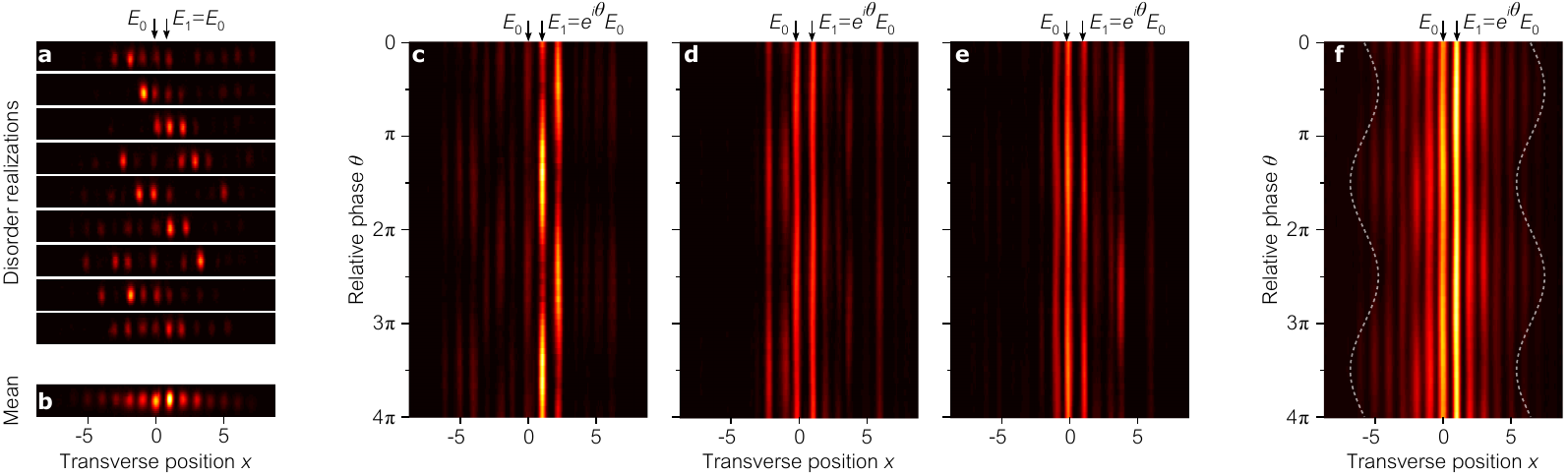}
	\caption{\label{Fig:IntensityPhase}  \textbf{The dependence of the output intensity across the lattice on the input relative phase. } \\ (a) Color plots depicting the intensity distributions for 9 different disordered lattice realizations captured by a CCD camera. Each plot represents a single realization when two input lattice sites illuminated with the relative phase $\theta\!=\!0$.  (b) An ensemble average obtained from 30 realizations. (c)-(e) Color plots depicting the output intensity distributions $I(x,\theta)$ while varying the input relative phase $\theta$ for three different disorder realizations. Each row is generated by integration along the $y$-direction of the CCD images (such as those in (a)). The three color plots are normalized to the \textit{same} peak value. The arrows at the top identify the input waveguides. (f) The ensemble average (30 realizations) of the intensity distribution $\langle I(x,\theta) \rangle $. The dashed white lines are guides to the eye highlighting the variation in the spatial offset of the mean intensity distribution with $\theta$.}
\end{figure*}

\section{Photonic thermalization gap}
Disorder-immune chiral symmetry has a critical impact on the statistics of light emerging from the lattice \cite{Kondakci2015b}. To characterize the strength of fluctuations in the optical field at lattice site $x$ and axial position $z$, we make use throughout of the normalized intensity correlation \cite{Goodman2000} $g_x^{(2)}\!(z)\!=\!\langle I_x^2(z)\rangle/\langle I_x(z) \rangle^2$. As such, coherent light is characterized by $g^{(2)}\!=\!1$ and incoherent (thermal) light by $g^{(2)}\!=\!2$. The ranges $1\!<\!g^{(2)}\!<2$ and $g^{(2)}\!>\!2$ delineate by convention sub-thermal and super-thermal light statistics, respectively.

When a single waveguide is excited with coherent light in a lattice with off-diagonal disorder, the output field in the same waveguide exhibits only super-thermal statistics even for small-sized lattices $2N+1\!\gtrsim\!15$. In fact, numerical and theoretical analyses \cite{Kondakci2015b} indicate that in the steady state $g^{(2)}\!\rightarrow\!3$ when $\Delta C\!\rightarrow\!0$, while $g^{(2)}\!\rightarrow\!2$ when $\Delta C\!\rightarrow\!1$ (solid black line in Fig.~\ref{Fig:numerical}(c)). In other words, a photonic thermalization gap opens up in this class of disordered lattices: the range of sub-thermal statistics is inaccessible. In disordered systems \textit{lacking} chiral symmetry, as in diagonally disordered lattices \cite{Anderson1958a} (dissimilar waveguides with identical couplings), the span of sub-thermal statistics \textit{is} accessible in the excitation waveguide while the super-thermal is not. In such lattices, the photonic thermalization gap is absent.

\section{Chiral-symmetry activation}
Observing this photonic thermalization gap in a lattice with off-diagonal disorder is subject, however, to first activating the chiral symmetry, which requires that the illumination satisfy a \textit{symmetric-excitation} condition, $|c_{n}|\!=\!|c_{-n}|$; that is, both modes in each chiral pair are excited with equal weights. This condition guarantees that the relative phase between the fields at \textit{any} two neighboring sites for all $z$ is always $\pm\pi/2$, which can be shown to produce only super-thermal statistics (Appendix). One example that satisfies this condition is that of single-site excitation described above. On the other hand, when the mode-excitation is \textit{a}symmetric, $|c_{n}|\!\neq\!|c_{-n}|$, then chiral symmetry remains dormant, or is de-activated, which may allow access to sub-thermal statistics \cite{Kondakci2015b}. The phases between adjacent lattice sites are no longer constrained and can take on arbitrary values. In fact, the field dynamics in a disordered lattice with chiral symmetry but broken excitation-symmetry can resemble that of a lattice lacking chiral symmetry altogether.

A simple field structure that enables tailoring the light statistics is that of two-site excitation. In our study, we excite neighboring waveguides at $x\!=\!0$ and $1$ coherently with equal amplitudes and relative phase $\theta$, $E_1(0)\!=\!e^{i\theta}E_0(0)$ (Fig.~\ref{Fig:numerical}(b)). In this case, $c_{\pm n}\!=\!\varphi_n(0)\!\pm\!e^{i\theta}\varphi_{n}(1)$, which satisfies the symmetric-excitation condition only when $\theta\!=\!\pm\pi/2$. Gradually increasing $\theta$ from 0 to $\pi/2$, thereby decreasing the violation of excitation-symmetry, reduces $g^{(2)}$ as depicted in Fig.~\ref{Fig:numerical}(c) (dashed lines). Excitation-symmetry is further confirmed directly through the modal decompositions shown in Fig.~\ref{Fig:numerical}(d).

\section{Light-Statistics Interferometry}
The output field amplitude at the center waveguide ($x\!=\!0$), $E_{0}(z)\!=\!E_{0,0}(z)\!+\!ie^{i\theta}E_{0,1}(z)$, receives contributions $E_{0,0}(z)\!=\!\!\sum_{n=1}^{N}\phi_n^2(0)\cos(b_n z)$ and $E_{0,1}(z)\!=\!\!\sum_{n=1}^{N}\phi_n(0)\phi_n(1) \sin(b_n z)$ from input sites 0 and 1, respectively, by virtue of the linearity of the system. The relative phase $\theta$ is imposed externally, while the $(\pi/2)$-phase is a consequence of chiral symmetry. Critically, this $(\pi/2)$-phase occurs in \textit{every realization}, such that the mean output intensity is
\begin{equation}\label{MeanIntensity}
\langle I_0 (z) \rangle = \langle I_{0,0}(z)\rangle + \langle I_{0,1}(z)\rangle - 2\sin \theta\langle E_{0,0}(z)E_{0,1}(z)\rangle, 
\end{equation}
where $I_{0,0}(z)\!=\!|E_{0,0}(z)|^2$ and $I_{0,1}(z)\!=\!|E_{0,1}(z)|^2$ (see Appendix for the general case of $x\!\neq\!0$). The last term in Eq.~\ref{MeanIntensity} vanishes in general at all output sites for large $z$, thereby rendering the mean intensity an incoherent sum of contributions $\langle I_{0,0}(z)\rangle$ and $\langle I_{0,1}(z)\rangle$ from the two input sites, which renders the output independent of $\theta$. The dependence of the output field on $\theta$ nevertheless remains prominent when examining $g^{(2)}$, which has the form
\begin{equation}\label{PhotonStatisticsInterferogram}
g^{(2)}_0(\theta)=\alpha-\beta\cos2\theta.
\end{equation}
The result is thus a \textit{light-statistics interferogram} (Fig.~\ref{Fig:Intro}(d)) with period \textit{half} that of the corresponding intensity interferogram from a typical deterministic interferometer (Fig.~\ref{Fig:Intro}(b)). Unlike intensity interferograms where the visibility captures the \textit{relative} swing in values, the \textit{absolute} values of $g^{(2)}$ are meaningful. The real, positive constants $\alpha$ and $\beta$ in Eq.~\ref{PhotonStatisticsInterferogram} are $\alpha=\eta_{0}g_{0,0}^{(2)}+\eta_{1}g_{0,1}^{(2)}+2\beta$ and $\beta\!=\!2\langle I_{0,0}I_{0,1}\rangle/\langle I_{0}\rangle^2$, where $g_{0,j}^{(2)}$ is the normalized intensity correlation at $x\!=\!0$ due to excitation at site $j$, $\eta_{j}\!=\!(\langle I_{0,j}\rangle/\langle I_{0}\rangle)^{2}$ is the squared fraction of input power contributed by site $j$, and $I_{0}\!=\!I_{0,0}+I_{0,1}$ is the total input power.

\section{Experiment} 
The photonic lattice in our experiment is a femtosecond-laser-written waveguide array \cite{Meany2015} consisting of 101 identical 49-mm-long waveguides with nearest-neighbor evanescent coupling. The numerical aperture of the waveguides is 0.06 and their average separation is 17~$\mu\mathrm{m}$ ($\bar{C}\!\approx\!0.71$~$\mathrm{cm}^{-1}$ at a wavelength of $\lambda\!=\!632$~nm). The waveguide separations are chosen such that the resulting coupling coefficients are uniformly distributed with $\Delta C\!\approx\!0.6$. Here, we note that since the relation between waveguide separation and coupling coefficient is not linear, the probability distribution of the waveguide separations is not uniform.  Ensemble averaging is produced by translating the  array in the transverse $x$ direction for 30 realizations \cite{Martin2011a} (Fig.~\ref{Fig:Setup}). A laser beam at $\lambda\!=\!632$~nm is split into two paths with controllable separation and relative phase, which are  coupled to pairs of adjacent waveguides through a $25\!\times$ microscope objective. A CCD camera records the output intensity distribution after magnification by a factor of 8, while the output at $x\!=\!0$ is concurrently imaged to a multimode fiber for a precise power measurement. Samples of the intensity distributions for different disorder realizations are given in Fig.~\ref{Fig:IntensityPhase}(a) for a relative phase $\theta\!=\!0$. Despite the disorder, the mean intensity distribution in the vicinity of the excitation waveguides $x\!=\!0$ and $1$ is stable (Fig.~\ref{Fig:IntensityPhase}(b)) as a result of Anderson localization.

We now proceed to exploit coherent control over chiral-symmetry-breaking to demonstrate \textit{light-statistics interferometry} -- deterministic tuning of the normalized intensity correlation $g^{(2)}$. To demonstrate continuous tuning of light statistics, we obtain the intensity distributions $I(x,\theta)$ for each disorder realization while varying $\theta$ in steps of $\pi/16$, three realizations of which are shown in Fig.~\ref{Fig:IntensityPhase}(c)-(e). Since the different realizations involve translating the array laterally with respect to the input beams, it is critical to identify a reliable reference for the phases across all the realizations. To address this challenge, we exploit a feature of the measurements acquired from individual realizations but which normally disappears after ensemble averaging, namely the interference term in Eq.~\ref{MeanIntensity}. This term enables identifying the relative values of $\theta$ between different realizations (modulo a phase of $\pi$; Appendix).

\begin{figure}[!thb]
	\centering\includegraphics[scale=1]{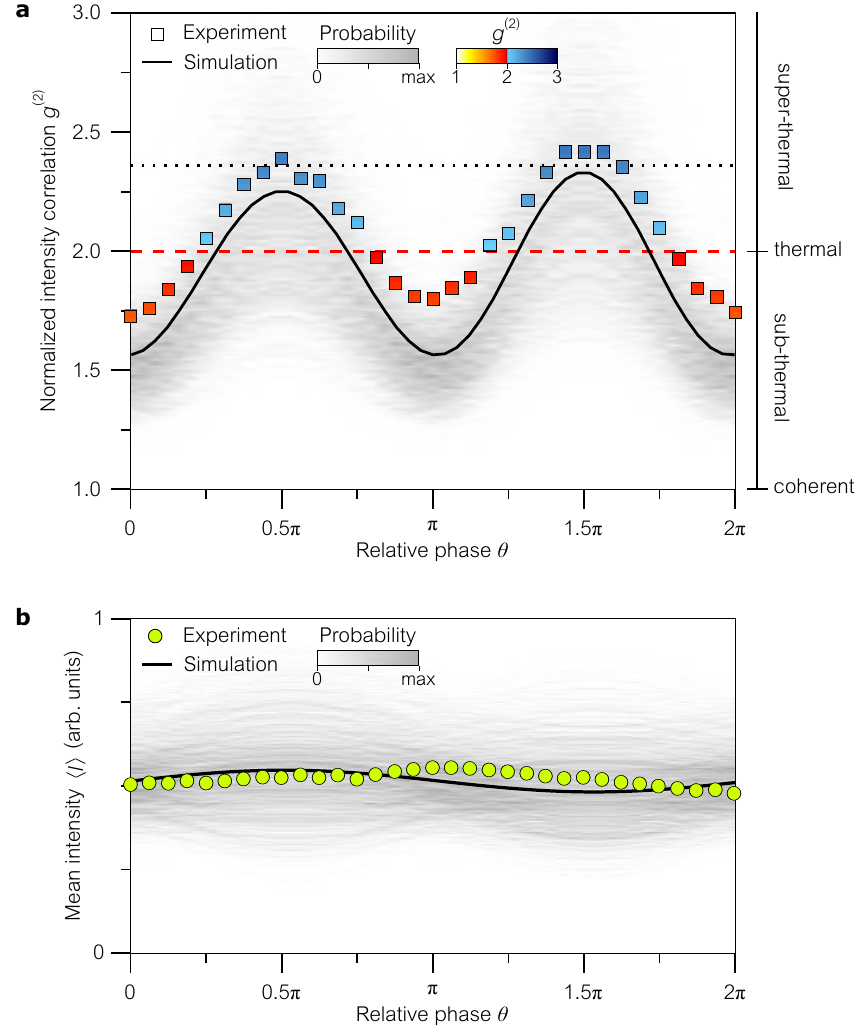}
	\caption{\label{Fig:Interferometry} \textbf{Light-statistics interferometry.} \\ (a) The normalized intensity correlation $g^{(2)}$ is deterministically tuned by varying the input relative phase $\theta$. Simulations (solid line) are in agreement with the data (squares). The gray shading is the calculated probability distribution of the expected $g^{(2)}$ values assuming a small ensemble size of 30 realizations (the size of the experimental ensemble), while the solid line is the average value of $g^{(2)}$ for an ensemble size of $10^{5}$ realizations. The red-dashed line corresponds to the edge of the photonic thermalization gap and separates the sub- and super-thermal regimes. The dotted line at $g^{(2)}\!=\!2.35$ is the value produced at the output when only one input lattice site is illuminated (and no tuning is available).  (b) The mean intensity as a function of $\theta$. The small-amplitude oscillation in the simulation (solid line) is due to the finite array length. The gray shading is the probability distribution of the mean intensity calculated for a small ensemble of 30 realizations, while the solid line was calculated for an ensemble of $10^5$ realizations.}
\end{figure}

We evaluate the normalized intensity correlation $g_0^{(2)}(\theta)$ using the intensity measurements collected by the multimode fiber, and present the experimental photon-statistics interferogram in Fig.~\ref{Fig:Interferometry}(a). We obtain a tuning-range of $g_0^{(2)}$ spanning $\approx 1.7$ to 2.4 -- from sub-thermal ($g^{(2)}\!<\!2$) to super-thermal ($g^{(2)}\!>\!2$). Numerical simulations for an ensemble size of $10^5$ are in good agreement with the measurements except for a small vertical offset in the value of $g_0^{(2)}$. The origin of this discrepancy can be traced to two effects: the finite size of the measurement ensemble and a mismatch between the excitation values at the two input waveguides (which is deduced from the unequal output intensities at $x\!=\!0$ and $1$, Fig.~\ref{Fig:IntensityPhase}(b)). We have simulated the probability distribution of $g_0^{(2)}$ for a small ensemble size (30 samples; shown in gray scale) and our experimental result falls within this region. The mean intensity remains approximately constant with $\theta$ throughout this procedure (Fig.~\ref{Fig:Interferometry}(b)). The remaining variation in the measured intensity is attributed solely to the finite array length.

\section{CONCLUSION}
We have developed a new interferometric methodology in which light statistics (quantified by $g^{(2)}$) is modulated deterministically -- while maintaining a fixed mean intensity -- by varying a relative phase between two coherent fields entering a finite disordered photonic network. In the process, we have confirmed the first observation of the predicted `photonic thermalization gap' in disordered lattices by virtue of their chiral symmetry\cite{Kondakci2015b} -- the disorder-immune feature that lays the foundation for coherent control of light statistics. By exploiting the thermalization gap associated with off-diagonal disorder, we coherently activate and de-activate the excitation-symmetry of the chiral-mode pairs to produce light whose statistics span the subthermal and superthermal regimes \textit{across} the edge of the thermalization gap.

Further modification of the input excitation can enable tuning the value of $g^{(2)}$ across \textit{all} the output lattice sites simultaneously. This requires illuminating the lattice with uniform intensity and phase differences of $\pm\pi/2$ between neighboring lattice sites \cite{Kondakci2015}. Our strategy can be extended to other on-chip implementations, such as coupled-resonator chains in which applied random voltages can modulate the couplings between resonators \cite{Mookherjea2014} to realize a versatile platform for dynamical control of light statistics in a compact device.

Our experiment poses a fundamental question: what classes of disordered systems permit tuning the output statistics via deterministic and coherent control over the excitation \textit{without} altering the system itself? Such systems generalize traditional interferometric paradigms to statistical quantities that are critical for energy transport. Because the principle behind coherent control of $g^{(2)}$ is the existence of a disorder-immune symmetry (the occurrence of chiral-mode pairs), one can ask whether the approach outlined here may be implemented in free space to tune the contrast of `chiral-like' speckled light. Such a tunable source could present a  powerful tool for  imaging through turbid media \cite{Saleh2011, Redding2012, Bertolotti2012, Bromberg2014, Katz2012, Matthews2014a, Zhou2014a}. Finally, our results pave the way to  deterministically tuning the  photon-number distributions \cite{Saleh1978} for low-intensity classical coherent light and non-classical light such as entangled photon pairs \cite{Abouraddy2012b, DiGiuseppe2013, Crespi2013,Gilead2015} or Fock states \cite{Mandel1995}.

\subsection*{ACKNOWLEDGMENTS}	
A.F.A. was supported by the U.S. Office of Naval Research (ONR) under contract N00014-14-1-0260. A.S. gratefully acknowledges financial support from the German Ministry of Education and Research (Center for Innovation Competence program, Grant No. 03Z1HN31).

\section*{APPENDIX}	
\appendix

\subsection{Data analysis} 
Since the inter-waveguide separations in a lattice with off-diagonal disorder are random, the separation between the two input beams must in turn be varied accordingly for optimal coupling as the array is translated along $x$. It is therefore critical to obtain a reliable reference for the relative phase $\theta$ across all the realizations. To explain our approach, we consider the same excitation scenario examined in the main text, namely $E_{0}(z\!=\!0)=e^{i\theta}E_{1}(0)$, but generalize the result to the output fields at $x\!\neq\!0$ and \textit{not} average over an ensemble. In general, the output field amplitude at $x$, 
$$E_{x}(z) = 
\begin{cases} 
i E_{x,0}(z) + e^{i\theta} E_{x,1}(z), \qquad x~\mathrm{odd}, \\ 
E_{x,0}(z) + ie^{i\theta}E_{x,1}(z), \qquad x~\mathrm{even},
\end{cases}$$
receives contributions 
$$E_{x,0}(z) = 
\begin{cases}
\sum_{n=1}^{N}\phi_n(x) \phi_n(0) \sin(b_n z), \qquad x~\mathrm{odd},  \\ 
\sum_{n=1}^{N}\phi_n(x) \phi_n(0) \cos(b_n z), \qquad x~\mathrm{even}, 
\end{cases} $$ 
and
$$E_{x,1}(z) = 
\begin{cases}
\sum_{n=1}^{N}\phi_n(x) \phi_n(1) \cos(b_n z), \qquad x~\mathrm{odd},  \\ 
\sum_{n=1}^{N}\phi_n(x) \phi_n(1) \sin(b_n z), \qquad x~\mathrm{even}, 
\end{cases} $$ 
from input sites 0 and 1, respectively. The output intensity is
\begin{equation}\label{Eq:SingleRealization}
I_x (z;\theta) = I_{x,0}(z)+ I_{x,1}(z) - 2p\sin\theta\sqrt{I_{x,0}(z)I_{x,1}(z)},
\end{equation}
where $I_{x,0}(z)\!=\!|E_{x,0}(z)|^2$,  $I_{x,1}(z)\!=\!|E_{x,1}(z)|^2$, and $p\!=\!\pm1$ varies randomly for different realizations. The third term in Eq.~\ref{Eq:SingleRealization} washes out after ensemble averaging, but is retained in individual realizations, such as those shown in Fig.~\ref{Fig:IntensityPhase}(c)-(e). What this interference term entails is that along \textit{each} waveguide in the individual realizations (not necessarily only $x\!=\!0$) an intensity interferogram emerges with $\theta$. The maxima and minima of these interferograms in \textit{all} waveguides and \textit{all} realizations can then be `lined up' to ensure that $\theta$ is calibrated with respect to the same origin. There remains the factor $p$ in Eq.~\ref{Eq:SingleRealization} which shifts the interferogram along $\theta$ by $\pi$ randomly from one realization to another, so that our phase reference is in actuality modulo-$\pi$. Although further symmetries in the individual realizations can be exploited to resolve this final ambiguity, we have found that it does not affect the value of $g^{(2)}$ resulting from averaging over the statistical ensemble.

\subsection{Chi-squared distribution}
In probability theory, the sum of the squares of $k$ independent random variables all of which have Gaussian probability distributions is a random variable characterized by a chi-squared probability distribution with $k$-degrees of freedom \cite{Papoulis1965}. For Gaussian distributions with zero mean and unity variance, the chi-squared distribution has a variance of $2k$ and mean of $k$. Consequently, the normalized intensity correlation is given by $$g^{(2)}=1+\frac{2}{k},~~k=1,2,\cdots.$$ When the excitation-symmetry condition is satisfied ($\theta\!=\!\pm\pi/2$) in a lattice with off-diagonal disorder, the phase constraint described in the main text dictates that the resulting field in any waveguide depends on a single random variable, $k=1$, and thus $g^{(2)}\!=\!3$ (super-thermal statistics). Alternatively, When the eigenmode-pairs are excited anti-symmetrically ($\theta\!=\!0$ or $\pi$), the field is complex and the real and imaginary components have identical probability density distributions. In this case, the intensity is the sum of the amplitude-squared of these two random variables, $k=2$, which entails that $g^{(2)}\!=\!2$ (thermal statistics)\cite{Goodman2000}.


\begin{thebibliography}{10}
\newcommand{\enquote}[1]{``#1''}

\bibitem{Born1999}
M.~Born and E.~Wolf, \emph{Principles of Optics} (Cambridge Univ. Press, 1999).

\bibitem{Goodman2000}
J.~W. Goodman, \emph{Statistical Optics} (John Wiley \& Sons, Inc., 2000).

\bibitem{Saleh2011}
B.~E.~A. Saleh, \emph{Introduction to Subsurface Imaging} (Cambridge Univ.
  Press, 2011).

\bibitem{Mosk2012}
A.~P. Mosk, A.~Lagendijk, G.~Lerosey, and M.~Fink, \enquote{Controlling waves
  in space and time for imaging and focusing in complex media,} Nature Photon.
  \textbf{6}, 283--292 (2012).

\bibitem{Vellekoop2010}
I.~M. Vellekoop, A.~Lagendijk, and A.~P. Mosk, \enquote{Exploiting disorder for
  perfect focusing,} Nature Photon. \textbf{4}, 320--322 (2010).

\bibitem{Judkewitz2013}
B.~Judkewitz, Y.~M. Wang, R.~Horstmeyer, A.~Mathy, and C.~Yang,
  \enquote{Speckle-scale focusing in the diffusive regime with time reversal of
  variance-encoded light ({TROVE}),} Nature Photon. \textbf{7}, 300--305
  (2013).

\bibitem{Nixon2013}
M.~Nixon, O.~Katz, E.~Small, Y.~Bromberg, A.~A. Friesem, Y.~Silberberg, and
  N.~Davidson, \enquote{Real-time wavefront shaping through scattering media by
  all-optical feedback,} Nature Photon. \textbf{7}, 919--924 (2013).

\bibitem{Zhou2014a}
E.~H. Zhou, H.~Ruan, C.~Yang, and B.~Judkewitz, \enquote{Focusing on moving
  targets through scattering samples,} Optica \textbf{1}, 227--232 (2014).

\bibitem{Lai2015}
P.~Lai, L.~Wang, J.~W. Tay, and L.~V. Wang, \enquote{Photoacoustically guided
  wavefront shaping for enhanced optical focusing in scattering media,} Nature
  Photon. \textbf{9}, 126--132 (2015).

\bibitem{Denault2013}
K.~A. Denault, M.~Cantore, S.~Nakamura, S.~P. DenBaars, and R.~Seshadri,
  \enquote{Efficient and stable laser-driven white lighting,} AIP Adv.
  \textbf{3}, 072107 (2013).

\bibitem{Redding2012}
B.~Redding, M.~A. Choma, and H.~Cao, \enquote{Speckle-free laser imaging using
  random laser illumination.} Nature Photon. \textbf{6}, 355--359 (2012).

\bibitem{Redding2015}
B.~Redding, A.~Cerjan, X.~Huang, M.~L. Lee, a.~D. Stone, M.~a. Choma, and
  H.~Cao, \enquote{Low spatial coherence electrically pumped semiconductor
  laser for speckle-free full-field imaging,} Proc. Natl. Acad. Sci. U. S. A.
  \textbf{112}, 1304--1309 (2015).

\bibitem{Bromberg2014}
Y.~Bromberg and H.~Cao, \enquote{Generating non-{R}ayleigh speckles with
  tailored intensity statistics,} Phys. Rev. Lett. \textbf{112}, 213904 (2014).

\bibitem{Gade1991}
R.~Gade and F.~Wegner, \enquote{The n=0 replica limit of {U(n)} and
  {U(n)}/{SO}(n) models,} Nucl. Phys. B \textbf{360}, 213--218 (1991).

\bibitem{Gade1993}
R.~Gade, \enquote{{A}nderson localization for sublattice models,} Nucl. Phys. B
  \textbf{398}, 499--515 (1993).

\bibitem{Evangelou2003}
S.~N. Evangelou and D.~E. Katsanos, \enquote{Spectral statistics in
  chiral-orthogonal disordered systems,} J. Phys. A \textbf{36}, 3237--3254
  (2003).

\bibitem{Lahini2011a}
Y.~Lahini, Y.~Bromberg, Y.~Shechtman, A.~Szameit, D.~N. Christodoulides,
  R.~Morandotti, and Y.~Silberberg, \enquote{{H}anbury {B}rown and {T}wiss
  correlations of {A}nderson localized waves,} Phys. Rev. A \textbf{84}, 041806
  (2011).

\bibitem{Kondakci2015b}
H.~E. Kondakci, A.~F. Abouraddy, and B.~E.~A. Saleh, \enquote{A photonic
  thermalization gap in disordered lattices,} Nature Phys. \textbf{11},
  930--935 (2015).

\bibitem{Soukoulis1981a}
C.~M. Soukoulis and E.~N. Economou, \enquote{Off-diagonal disorder in
  one-dimensional systems,} Phys. Rev. B \textbf{24}, 5698--5702 (1981).

\bibitem{Szameit2010}
A.~Szameit, Y.~V. Kartashov, P.~Zeil, F.~Dreisow, M.~Heinrich, R.~Keil,
  S.~Nolte, A.~T{\"{u}}nnermann, V.~A. Vysloukh, and L.~Torner, \enquote{Wave
  localization at the boundary of disordered photonic lattices.} Opt. Lett.
  \textbf{35}, 1172--1174 (2010).

\bibitem{Martin2011a}
L.~Martin, G.~{Di Giuseppe}, A.~Perez-Leija, R.~Keil, F.~Dreisow, M.~Heinrich,
  S.~Nolte, A.~Szameit, A.~F. Abouraddy, D.~N. Christodoulides, and B.~E.~A.
  Saleh, \enquote{{A}nderson localization in optical waveguide arrays with
  off-diagonal coupling disorder.} Opt. Express \textbf{19}, 13636--13646
  (2011).

\bibitem{Christodoulides2003a}
D.~N. Christodoulides, F.~Lederer, and Y.~Silberberg, \enquote{Discretizing
  light behaviour in linear and nonlinear waveguide lattices.} Nature
  \textbf{424}, 817--823 (2003).

\bibitem{Schwartz2007a}
T.~Schwartz, G.~Bartal, S.~Fishman, and M.~Segev, \enquote{Transport and
  {A}nderson localization in disordered two-dimensional photonic lattices.}
  Nature \textbf{446}, 52--55 (2007).

\bibitem{Lahini2008a}
Y.~Lahini, A.~Avidan, F.~Pozzi, M.~Sorel, R.~Morandotti, D.~N. Christodoulides,
  and Y.~Silberberg, \enquote{{A}nderson localization and nonlinearity in
  one-dimensional disordered photonic lattices,} Phys. Rev. Lett. \textbf{100},
  013906 (2008).

\bibitem{Segev2013}
M.~Segev, Y.~Silberberg, and D.~N. Christodoulides, \enquote{{A}nderson
  localization of light,} Nature Photon. \textbf{7}, 197--204 (2013).

\bibitem{Anderson1958a}
P.~W. Anderson, \enquote{Absence of diffusion in certain random lattices,}
  Phys. Rev. \textbf{109}, 1492--1505 (1958).

\bibitem{Meany2015}
T.~Meany, M.~Gr{\"{a}}fe, R.~Heilmann, A.~Perez-Leija, S.~Gross, M.~J. Steel,
  M.~J. Withford, and A.~Szameit, \enquote{Laser written circuits for quantum
  photonics,} Laser Photon. Rev. \textbf{9}, 363--384 (2015).

\bibitem{Kondakci2015}
H.~E. Kondakci, A.~F. Abouraddy, and B.~E.~A. Saleh, \enquote{Discrete
  {A}nderson speckle,} Optica \textbf{2}, 201--209 (2015).

\bibitem{Mookherjea2014}
S.~Mookherjea, J.~R. Ong, X.~Luo, and L.~Guo-Qiang, \enquote{Electronic control
  of optical {A}nderson localization modes.} Nature Nanotech. \textbf{9},
  365--371 (2014).

\bibitem{Bertolotti2012}
J.~Bertolotti, E.~G.~. van Putten, C.~Blum, A.~Lagendijk, W.~L. Vos, and A.~P.
  Mosk, \enquote{Non-invasive imaging through opaque scattering layers.} Nature
  \textbf{491}, 232--234 (2012).

\bibitem{Katz2012}
O.~Katz, E.~Small, and Y.~Silberberg, \enquote{Looking around corners and
  through thin turbid layers in real time with scattered incoherent light,}
  Nature Photon. \textbf{6}, 549--553 (2012).

\bibitem{Matthews2014a}
T.~E. Matthews, M.~Medina, J.~R. Maher, H.~Levinson, W.~J. Brown, and A.~Wax,
  \enquote{Deep tissue imaging using spectroscopic analysis of multiply
  scattered light,} Optica \textbf{1}, 105 (2014).

\bibitem{Saleh1978}
B.~E.~A. Saleh, \emph{Photoelectron Statistics} (Springer, 1978).

\bibitem{Abouraddy2012b}
A.~F. Abouraddy, G.~{Di Giuseppe}, D.~N. Christodoulides, and B.~E.~A. Saleh,
  \enquote{{A}nderson localization and colocalization of spatially entangled
  photons,} Phys. Rev. A \textbf{86}, 040302 (2012).

\bibitem{DiGiuseppe2013}
G.~{Di Giuseppe}, L.~Martin, A.~Perez-Leija, R.~Keil, F.~Dreisow, S.~Nolte,
  A.~Szameit, A.~F. Abouraddy, D.~N. Christodoulides, and B.~E.~A. Saleh,
  \enquote{{E}instein-{P}odolsky-{R}osen spatial entanglement in ordered and
  {A}nderson photonic lattices,} Phys. Rev. Lett. \textbf{110}, 150503 (2013).

\bibitem{Crespi2013}
A.~Crespi, R.~Osellame, R.~Ramponi, V.~Giovannetti, R.~Fazio, L.~Sansoni,
  F.~{De Nicola}, F.~Sciarrino, and P.~Mataloni, \enquote{{A}nderson
  localization of entangled photons in an integrated quantum walk,} Nature
  Photon. \textbf{7}, 322--328 (2013).

\bibitem{Gilead2015}
Y.~Gilead, M.~Verbin, and Y.~Silberberg, \enquote{Ensemble-averaged quantum
  correlations between path-entangled photons undergoing {A}nderson
  localization,} Phys. Rev. Lett. \textbf{115}, 133602 (2015).

\bibitem{Mandel1995}
L.~Mandel and E.~Wolf, \emph{Optical Coherence and Quantum Optics} (Cambridge
  Univ. Press, 1995).

\bibitem{Papoulis1965}
A.~Papoulis, \emph{Probability, Random Variables and Stochastic Processes}
  (McGraw-Hill Book Comp., 1965).

\end{thebibliography}
\end{document}